\documentstyle[12pt,epsfig,a4]{article} 
\newcommand{\vs}{\vspace{-0.25cm}}
\begin{document} 

\begin{center}
{\Large{\bf Radiative corrections to pion-nucleus bremsstrahlung}}
\bigskip

N. Kaiser and J.M. Friedrich\\
\medskip
{\small Physik-Department, Technische Universit\"{a}t M\"{u}nchen,
    D-85747 Garching, Germany}
\end{center}
\medskip
\begin{abstract}
We calculate the one-photon loop radiative corrections to virtual pion
Compton scattering $\pi^- \gamma^* \to \pi^- \gamma$, that subprocess which 
determines in the one-photon exchange approximation the pion-nucleus 
bremsstrahlung reaction $\pi^- Z\to \pi^- Z \gamma$. Ultraviolet and infrared 
divergencies of the loop integrals are both treated by dimensional 
regularization. Analytical expressions for the ${\cal O}(\alpha)$ corrections 
to the virtual Compton scattering amplitudes, $A(s,u,Q)$ and $B(s,u,Q)$, are 
derived with their full dependence on the (small) photon virtuality $Q$ from 9 
classes of contributing one-loop diagrams. Infrared finiteness of these 
virtual radiative corrections is achieved (in the standard way) by including 
soft photon radiation below an energy cut-off $\lambda$. In the region of low 
$\pi^- \gamma$ center-of-mass energies, where the pion-nucleus bremsstrahlung 
process is used to extract the pion electric and magnetic polarizabilities, we 
find radiative corrections up to about $-3\%$ for $\lambda=5\,$MeV. 
Furthermore, we extend our calculation of the radiative corrections to virtual 
pion Compton scattering $\pi^-\gamma^* \to \pi^- \gamma$ by including the 
leading pion structure effect in form of the polarizability difference
$\alpha_\pi - \beta_\pi$. Our analytical results are particularly relevant for 
analyzing the data of the COMPASS experiment at CERN which aims at measuring 
the pion electric and magnetic polarizabilities with high statistics using the 
Primakoff effect.    
\end{abstract}

\bigskip
\bigskip

\noindent

\section{Introduction and summary}
At present there is much interest in a precise experimental determination of 
the pion electric and magnetic polarizabilities, $\alpha_\pi$ and $\beta_\pi$. 
Within the framework of current algebra \cite{terentev} it has been shown 
(long ago) that the polarizability difference $\alpha_\pi-\beta_\pi$ of the 
charged pion is directly related to the axial-vector-to-vector form factor 
ratio $h_A/h_V \simeq 0.44$ measured in the radiative pion decay $\pi^+\to e^+ 
\nu_e \gamma$ \cite{frlez}. At leading (nontrivial) order the result of chiral 
perturbation theory \cite{doho}, $\alpha_\pi- \beta_\pi =\alpha(\bar l_6- \bar 
l_5)/(24 \pi^2 f_\pi^2 m_\pi)$, is of course the same after identifying the 
low-energy constant as $\bar l_6-\bar l_5 = 6h_A/h_V$. Recently, the systematic 
corrections to this current algebra result have been worked out in 
refs.\cite{buergi,gasser} by performing a full two-loop calculation of (real) 
pion Compton scattering in chiral perturbation theory. The outcome of that 
extensive analysis is that altogether the higher order corrections are rather 
small and the value $\alpha_\pi-\beta_\pi=(5.7\pm 1.0)\cdot 10^{-4}\,$fm$^3$ 
\cite{gasser} for the pion polarizability difference stands now as a firm 
prediction of the (chiral-invariant) theory. The non-vanishing value 
$\alpha_\pi +\beta_\pi= (0.16\pm 0.1) \cdot 10^{-4}\,$fm$^3$ \cite{gasser} for
the pion polarizability sum (obtained also at two-loop order) is presumably 
much too small to cause an observable effect in low-energy pion Compton
scattering.  

However, the chiral prediction $\alpha_\pi- \beta_\pi=(5.7\pm 1.0)\cdot 
10^{-4}\,$fm$^3$ is in conflict with the existing experimental determinations 
of $\alpha_\pi- \beta_\pi=(15.6\pm 7.8) \cdot 10^{-4}\,$fm$^3$ from Serpukhov 
\cite{serpukhov} and $\alpha_\pi-\beta_\pi =(11.6\pm 3.4)\cdot 10^{-4}\,$fm$^3$ 
from Mainz \cite{mainz}, which amount to values more than twice as large. 
Certainly, these existing experimental determinations of $\alpha_\pi-\beta_\pi$ 
raise doubts about their correctness since they violate the chiral low-energy 
theorem notably by a factor 2. 

In that contradictory situation it is promising that the ongoing COMPASS 
\cite{compass} experiment at CERN aims at measuring the pion polarizabilities 
with high statistics using the Primakoff effect. The scattering of high-energy 
negative pions in the Coulomb field of a heavy nucleus (of charge $Z$) gives 
access to cross sections for $\pi^- \gamma$ reactions through the equivalent
photons method \cite{pomer}. In practice, one preferentially analyzes the 
spectrum of bremsstrahlung photons produced in the reaction $\pi^- Z\to\pi^- 
Z\gamma$ in the so-called Coulomb peak. This kinematical regime is 
characterized by very small momentum transfers to the nuclear target such that 
virtual pion Compton scattering $\pi^- \gamma^* \to\pi^- \gamma$ occurs as the 
dominant subprocess (in the one-photon exchange approximation). The deviations 
of the measured spectra (at low $\pi^-\gamma$ center-of-mass energies) from 
those of a point-like pion are then attributed to the pion structure as 
represented by its electric and magnetic polarizability \cite{serpukhov} 
(taking often $\alpha_\pi+\beta_\pi=0$ as a constraint). It should be stressed 
here that the systematic treatment of virtual pion Compton scattering $\pi^- 
\gamma^* \to \pi^- \gamma$ in chiral perturbation theory yields at the same 
order as the polarizability difference $\alpha_\pi-\beta_\pi$ a further 
pion-structure effect in form of a unique pion-loop correction 
\cite{unkmeir,radcor} (interpretable as photon scattering off the ``pion-cloud 
around the pion''). In the case of real pion Compton scattering $\pi^- \gamma
\to \pi^- \gamma$ this loop correction compensates partly the effects from the
pion polarizability difference \cite{picross}. A minimal requirement for
improving future analyses of pion-nucleus bremsstrahlung $\pi^- Z\to\pi^-
Z\gamma$ is therefore to include the loop correction predicted by chiral
perturbation theory. The relevant analytical formulas for doing this have been
written down in the appendix of  ref.\cite{radcor}. 

At the required level of accuracy it is also necessary to include higher order
electromagnetic corrections to the pion-nucleus bremsstrahlung reaction $\pi^-
Z\to\pi^- Z\gamma$. Such radiative corrections have been considered some time
ago in the works of Akhundov et al. \cite{akhundov} and they have been 
implemented into the numerical program called RCFORGV. Unfortunately, no 
accessible sources to the underlying analytical expressions for the photon-loop
amplitudes (which are necessary for an independent implementation into new data 
analyses) have been given by these authors. The purpose of the present work is 
to fill this gap. We will re-evaluate the one-photon loop radiative corrections 
to virtual pion Compton scattering $\pi^- \gamma^* \to\pi^- \gamma$ and present 
for each class of diagrams the corresponding analytical expressions. The
calculation gets considerably simplified by considering the T-matrix of
$\pi^- \gamma^* \to\pi^- \gamma$ in the laboratory frame and exploiting the
circumstance that the time-like polarization vector $(1,\vec 0\,)$ of the
virtual photon is orthogonal to the space-like polarization vector $(0,\vec
\epsilon\,)$ of the real bremsstrahlung photon. Incorporating furthermore
crossing-symmetry, one finds that only two dimensionless amplitudes, $A(s,u,Q)$
and $B(s,u,Q)$, depending on three dimensionless Lorentz-invariant
kinematical variables, need to be specified.   

Our paper is organized as follows. In section\,2, we review the cross section
for pion-nucleus bremsstrahlung (in the one-photon exchange approximation)
and define the T-matrix of virtual pion Compton scattering $\pi^- \gamma^* 
\to\pi^- \gamma$ in the laboratory frame together with the tree-level Born
contributions. Section\,3 is devoted to the calculation of the one-photon loop 
radiative corrections. We present analytical expressions for the amplitudes 
$A(s,u,Q)$ and $B(s,u,Q)$ of order $\alpha$ as they emerge from 9 classes of 
one-photon loop diagrams. Dimensional regularization is used to treat both 
ultraviolet and infrared divergencies of the loop integrals. While the 
ultraviolet divergencies drop out in the renormalizable scalar quantum 
electrodynamics at work, the infrared divergencies get removed (in the
standard  way) by 
including soft photon radiation below an energy cut-off $\lambda$. We  
evaluate in section\,4 the resulting finite (real) radiative correction factor 
$\delta_{\rm real}^{\rm (cm)}$ with a small energy cut-off $\lambda$ introduced in 
the $\pi^-\gamma$ center-of-mass frame. Numerical results are presented for 
this part of the radiative corrections in the region of low $\pi^-\gamma$ 
invariant masses, where pion-nucleus bremsstrahlung is used to extract the 
pion polarizabilities. The detailed evaluation of the finite (virtual) 
radiative correction factor is not possible at this stage, since it requires 
specification and implementation of the actual experimental conditions (such 
as pion beam energy, constraints on the detectable energy and angular ranges,
etc.). In section\,5, we go beyond the works of Akhundov et al. \cite{akhundov}
and extend the calculation of the radiative corrections to virtual pion
Compton scattering  $\pi^- \gamma^* \to\pi^- \gamma$ by including the leading
pion structure effect (at low energies) in form of the polarizability
difference $\alpha_\pi- \beta_\pi$. In an effective field theory approach this 
feature can be elegantly handled by a suitable (gauge-invariant) two-photon 
contact-vertex. In the appendix, we study some two-photon exchange processes
specific  to scalar quantum electrodynamics which turn out to be negligibly 
small. 

Our analytical results for the radiative corrections to virtual pion Compton 
scattering can be utilized for analyzing the data of the COMPASS experiment at 
CERN which aims at measuring the pion polarizabilities via high-energy
pion-nucleus bremsstrahlung  $\pi^-Z\to\pi^- Z\gamma$.   
  
\section{Cross section for pion-nucleus bremsstrahlung} 
In order to keep this paper self-contained, we start with reviewing the cross
section for pion-nucleus bremsstrahlung in the one-photon exchange
approximation. Consider the process, $\pi^-(\vec p_1) + Z_{\rm rest} \to 
\pi^-(\vec p_2)+ \gamma(\vec k, \vec \epsilon\,) + Z_{\rm recoil}$, where a
virtual photon transfers the (small) momentum $\vec q =\vec p_2+\vec k-\vec p_1$
from the nucleus (of charge $Z$) at rest to the (high-energetic) charged pion. 
The extremely small recoil energy $-q_0 = \vec q\,^2/2M_{\rm nucl} \approx 0 $
of the nucleus can be (safely) neglected. The corresponding fivefold 
differential cross section in the laboratory  frame reads \cite{itzykson}: 
\begin{equation} {d^5 \sigma \over d\omega d\Omega_\gamma d\Omega_\pi} = {Z^2
    \alpha^3 \omega |\vec p_2| \over \pi^2 |\vec p_1| |\vec q\,|^4} \,H\,, 
\end{equation}
with $\alpha=1/137.036$ the fine-structure constant and $H$ a squared 
amplitude summed over the final state (real) photon polarizations 
$(\vec \epsilon \cdot \vec k = 0)$: 
\begin{equation} H =|\tilde A\,\vec p_1\sin \theta_1|^2+|\tilde B\,\vec  p_2\sin
\theta_2|^2 + 2{\rm Re}(\tilde A \tilde B^*) |\vec p_1| |\vec p_2| \sin \theta_1 
\sin \theta_2 \cos \phi\,.\end{equation} 
It has been derived from a T-matrix for the virtual Compton scattering
subprocess, $\pi^-\gamma^*_0 \to \pi^-\gamma$, of the form $T_{\rm lab} =- 8\pi 
\alpha (\tilde A \,\vec \epsilon\,^* \cdot \vec p_1+ \tilde B\,\vec\epsilon\,^* 
\cdot \vec p_2)$ with $\tilde A $ and $\tilde B$ two (in general complex)  
amplitudes. Since the electromagnetic interaction with the heavy nucleus  
involves only the charge density, the polarization four-vector of the virtual
(Coulomb) photon has only a time-component: $(1, \vec 0\,)$. The quantities
$\theta_1$ and $\theta_2$ in eq.(2) are the angles between the momentum vectors 
$\vec k$ and $\vec p_1$ and between $\vec k$ and $\vec p_2$, respectively. 
Moreover, $\phi$ is the 
angle between the two planes spanned by $(\vec k,\vec p_1)$ and $(\vec k,
\vec p_2)$. The photon energy is denoted by $\omega=|\vec k\,| = E_1-E_2$,
where $E_1 = \sqrt{|\vec p_1|^{2} + m_\pi^2}$ and $E_2 = \sqrt{|\vec p_2|^{2}  + 
m_\pi^2}$ are the initial and final pion energies in the laboratory frame.

Taking into account crossing-symmetry, the T-matrix for virtual pion Compton 
scattering, $\pi^-(\vec p_1)+\gamma^*_0(\vec q\,) \to \pi^-(\vec p_2)+\gamma(
\vec k, \vec  \epsilon\,)$, in the laboratory frame can be further decomposed
as: 
\begin{equation} T_{\rm lab}={8\pi \alpha \over m_\pi^2}\bigg\{ \vec \epsilon^{\,\,*} 
\cdot\vec p_1 \Big[ E_1 \, A(s,u,Q)+ E_2 \, B(s,u,Q)\Big]+ \vec \epsilon^{\,\,*}
\cdot \vec p_2 \Big[ E_1 \, B(u,s,Q)+ E_2 \, A(u,s,Q)\Big]\bigg\} \,.
\end{equation}
Here, we have introduced the (convenient) dimensionless Mandelstam variables: 
$s=(p_2+k)^2/m_\pi^2$, $u= (p_1-k)^2/m_\pi^2$, $t= (p_1-p_2)^2/m_\pi^2$ and $Q= 
-q^2/m_\pi^2$ (given by Lorentz-squares of momentum four-vectors), which obey  
the constraint $s+t+u=2-Q$. In the physical region one has: $s>1$, $u<1$, 
$t<0$ and $Q>0$.
The less obvious condition $u-1<0$ can be deduced by considering its left hand 
side in the $\pi^-\gamma$ center-of-mass frame. Due to the prefactor $8\pi
\alpha/m_\pi^2$ in eq.(3) the two independent amplitudes $A(s,u,Q)$ and 
$B(s,u,Q)$ become also dimensionless quantities. The crossing symmetry of the 
T-matrix, which corresponds to an invariance under the transformation $(E_1, 
\vec p_1) \leftrightarrow (-E_2, -\vec p_2)$ and $s \leftrightarrow u$, is 
manifest according to the decomposition written in eq.(3).   

\begin{figure}
\begin{center}
\includegraphics[scale=0.95,clip]{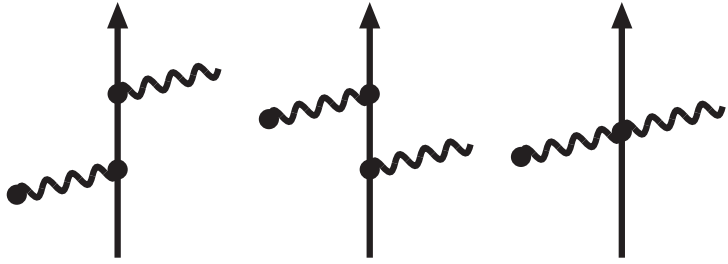}
\end{center}
\vspace{-.8cm}
\caption{Tree diagrams of scalar QED for virtual pion Compton scattering. The 
virtual photon on the left hand side couples to heavy nucleus.}
\end{figure}
The advantage of working with dimensionless amplitudes and variables shows up
already when calculating the Born terms for a point-like pion. The (three) 
tree diagrams of scalar quantum electrodynamics shown in Fig.\,1 lead to the
following extremely simple expressions:
\begin{equation} A(s,u)^{(\rm tree)} = 0\,, \qquad B(s,u)^{(\rm tree)} = 
{2\over u-1} \,,\end{equation} 
where we suppress from now on the dependence of $A(s,u,Q)$ and $B(s,u,Q)$ on 
the third variable $Q$. Note that the right diagram in Fig.\,1 involving the 
two-photon contact-vertex of scalar QED vanishes, due to the orthogonality of 
the time-like virtual photon and the space-like real photon polarizations, 
$\epsilon_0 =0$. The $s$-channel pole term (generated by the left diagram in 
Fig.\,1) enters into the T-matrix through the (crossed) amplitude $B(u,s)^{(\rm  
tree)}= 2/(s-1)$. The contributions of the pion polarizabilities 
(predominantly their difference $\alpha_\pi-\beta_\pi\simeq -2\beta_\pi$) to
the  amplitudes $A(s,u)$ and $B(s,u)$ can be found in eq.(27).     

\section{Evaluation of one-photon loop diagrams} 
In this section, we present analytical results for the one-photon loop 
radiative corrections of order $\alpha$ to the invariant amplitudes $A(s,u)$ 
and $B(s,u)$ of virtual pion Compton scattering. We use the method of 
dimensional regularization to treat both ultraviolet and infrared divergencies
(where the latter are caused by the masslessness of the photon). The method 
consists in calculating loop integrals in $d$ spacetime dimensions and 
expanding the results around $d=4$. Divergent pieces of one-loop integrals 
generically show up in form of the composite constant:
\begin{equation} \xi = {1\over d-4} +{1\over2} (\gamma_E-\ln 4\pi) +\ln{m_\pi
\over \mu}\,, \end{equation}
containing a simple pole at $d=4$. In addition, $\gamma_E = 0.5772\dots$ is the 
Euler-Mascheroni number and $\mu$ an arbitrary mass scale introduced in 
dimensional regularization in order to keep the mass dimension of the loop 
integrals independent of $d$. Ultraviolet (UV) and infrared (IR) divergencies 
are distinguished by the feature of whether the condition for convergence of 
the $d$-dimensional integral is $d<4$ or $d>4$. We discriminate them in the 
notation by putting appropriate subscripts, i.e. $\xi_{UV}$ and $\xi_{IR}$. The 
basic aspects of mass $(m_\pi)$ and coupling constant $(e)$ renormalization in
scalar QED have been discussed in section 2 of ref.\cite{radcor}.

The gauge condition $\epsilon_0=0$ eliminates those (five) loop diagrams where
both the incoming virtual photon and the outgoing real photon are connected by 
the two-photon contact-vertex of scalar QED (in first or second order). One can 
also drop the diagrams with a tadpole-type self-energy insertion (the loop 
consisting of a single closed photon line) since it is set to zero in 
dimensional regularization. Thus we are left with 24 one-photon loop diagrams 
which we group into the (separately crossing-symmetric) classes I-IX, shown 
in Figs.\,2-10. Due to an increasing number of internal pion propagators their 
evaluation rises in complexity. In order to simplify all calculations, we 
employ the Feynman gauge, where the photon propagator is directly proportional 
to  the Minkowski metric tensor $g^{\mu\nu}$. We can now enumerate the
analytical expressions for $A(s,u)$ and $B(s,u)$ as they emerge from the 9 
classes of contributing one-photon loop diagrams. 

\medskip

\begin{figure}
\begin{center}
\includegraphics[scale=0.95,clip]{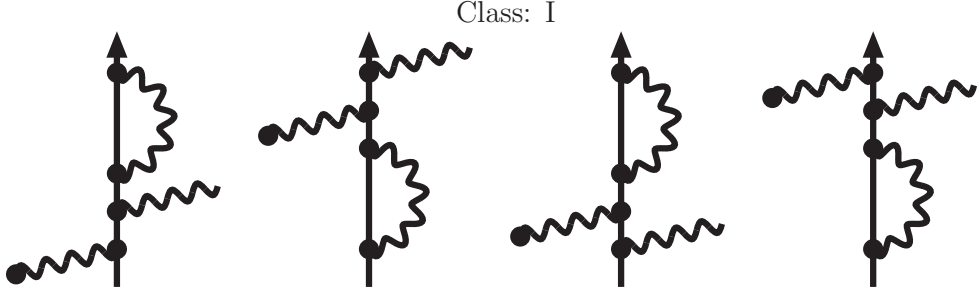}
\end{center}
\vspace{-.8cm}
\caption{One-photon loop diagrams of class I. The virtual photon comes in from 
the left.}
\end{figure}

\noindent
\underline{Class I}: For this class of diagrams the tree level amplitude in 
eq.(4) gets multiplied with the wavefunction renormalization factor of the pion:
\begin{equation} B(s,u)^{(\rm I)} = {2\alpha \over\pi(u-1)} \Big( \xi_{IR}  -\xi_{UV}
\Big)\,.\end{equation}

\begin{figure}
\begin{center}
\includegraphics[scale=0.95,clip]{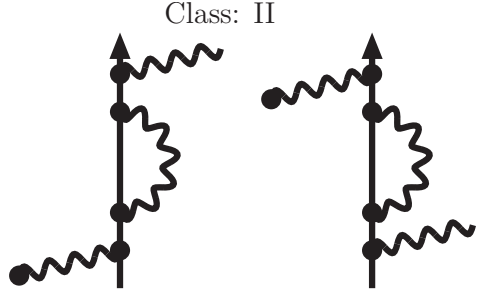}
\end{center}
\vspace{-.8cm}
\caption{One-photon loop diagrams of class II.}
\end{figure}

\noindent
\underline{Class II}: These diagrams involve the once-subtracted (off-shell) 
self-energy of the pion:
\begin{equation} B(s,u)^{(\rm II)} = {\alpha \over\pi(u-1)}\bigg[-2\xi_{UV}+2  
-{u+1 \over u} \ln(1-u) \bigg]\,.\end{equation}

\begin{figure}
\begin{center}
\includegraphics[scale=0.95,clip]{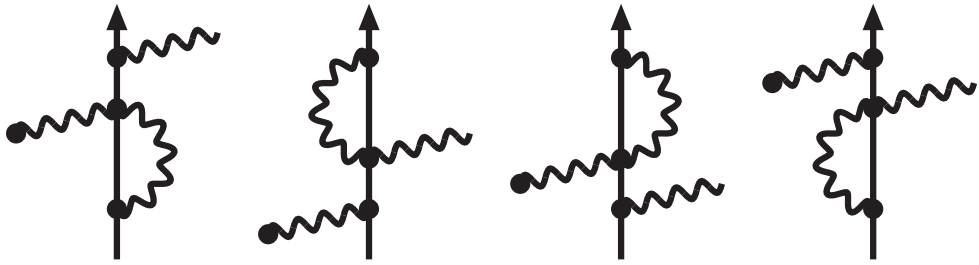}
\end{center}
\vspace{-.8cm}
\caption{One-photon loop diagrams of class III.}
\end{figure}

\noindent
\underline{Class III}: These diagrams generate a constant vertex correction 
factor:
\begin{equation} B(s,u)^{(\rm III)} = {\alpha \over\pi(u-1)}\bigg[3\xi_{UV}-{7\over 
2} \bigg]\,.\end{equation}

\begin{figure}
\begin{center}
\includegraphics[scale=0.95,clip]{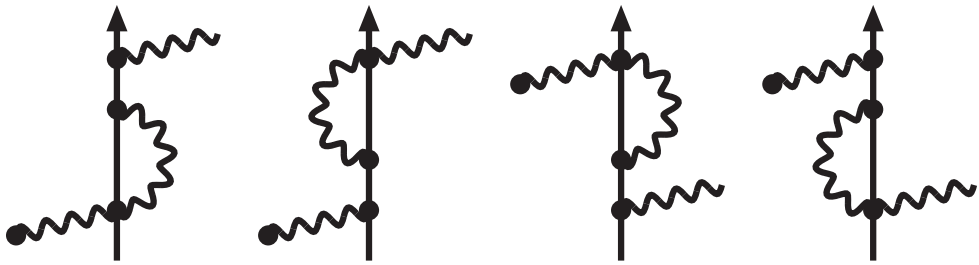}
\end{center}
\vspace{-.8cm}
\caption{One-photon loop diagrams of class IV.}
\end{figure}

\noindent
\underline{Class IV}: These diagrams  generate a $u$-dependent vertex 
correction factor:
\begin{equation} B(s,u)^{(\rm IV)} = {\alpha \over\pi(u-1)}\bigg[3\xi_{UV}-3  
-{1\over 2u}+{u-1 \over 2u^2}(3u+1) \ln(1-u) \bigg]\,.\end{equation}

\begin{figure}
\begin{center}
\includegraphics[scale=0.95,clip]{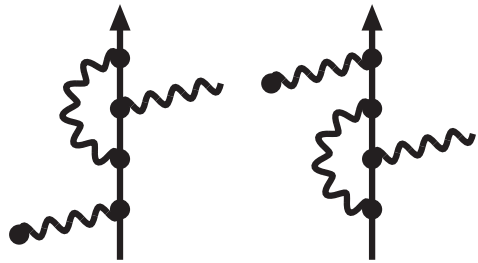}
\end{center}
\vspace{-.8cm}
\caption{One-photon loop diagrams of class V.}
\end{figure}

\noindent
\underline{Class V}: These diagrams  generate also a $u$-dependent vertex 
correction factor:
\begin{equation} B(s,u)^{(\rm V)} = {\alpha \over4\pi(u-1)} \bigg[-4\xi_{UV}+5
+{1\over u}+{u^2+6u+1 \over u^2} \ln(1-u) \bigg]\,.\end{equation}

\begin{figure}
\begin{center}
\includegraphics[scale=0.95,clip]{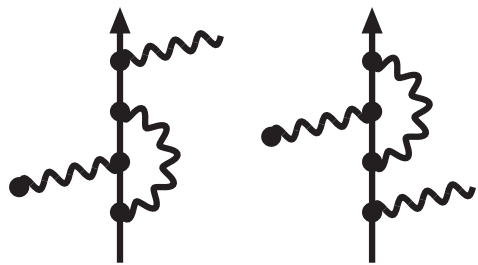}
\end{center}
\vspace{-.8cm}
\caption{One-photon loop diagrams of class VI.}
\end{figure}

\noindent
\underline{Class VI}: These diagrams  generate a $u$- and $Q$-dependent vertex 
correction factor:
\begin{eqnarray} B(s,u)^{(\rm VI)} &=&  {\alpha \over4\pi(u-1)}\bigg\{-4\xi_{UV}+5
+{1\over u}+{1-u^2 \over u^2}\ln(1-u) \nonumber \\ &&\qquad \qquad\quad -2(u+3+
2Q) \int_0^1\!\!dx\!\int_0^1\!\!dy\,\,{1-y\over K(u,Q)}\,\bigg\}\,,\end{eqnarray}
with the cubic polynomial $K(u,Q) = y+(1-u)x(1-y)+Q x(1-x)y$ in two Feynman
parameters $x,y$. Note that in the physical region $u<1, Q>0$ the denominator 
$K(u,Q)>0$ is strictly positive and therefore the integral in eq.(11) is 
well-defined and can be performed easily numerically. Interestingly, for the 
classes I-VI considered up to here, the contributions to the amplitude $A(s,u)$ 
vanish identically in each case.

\begin{figure}
\begin{center}
\includegraphics[scale=0.95,clip]{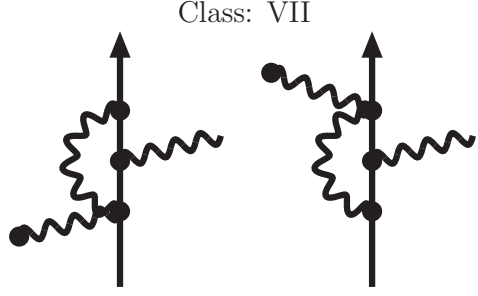}
\end{center}
\vspace{-.8cm}
\caption{One-photon loop diagrams of class VII.}
\end{figure}

\noindent
\underline{Class VII}: These irreducible diagrams give:
\begin{equation} A(s,u)^{(\rm VII)} = {\alpha \over4\pi(u-1)} \bigg\{{5-3u\over u}
\ln(1-u)-2 +{1 \over u-1}\bigg[ 2 {\rm Li}_2(u)-{\pi^2 \over 3}\bigg]\bigg\}
\,,\end{equation}
\begin{equation} B(s,u)^{(\rm VII)}= {\alpha \over4\pi(u-1)} \bigg\{{1+u \over u} +
  {1-3u\over u^2} \ln(1-u)+{1 \over u-1}\bigg[{\pi^2 \over 3}- 2 {\rm Li}_2(u)
\bigg]\bigg\} \,,\end{equation}
where Li$_2(u) = \sum_{n=1}^\infty n^{-2} u^n = u \int_1^\infty dx [x(x-u)]^{-1}\ln x$ 
denotes the conventional dilogarithmic function. 
\begin{figure}

\begin{center}
\includegraphics[scale=0.95,clip]{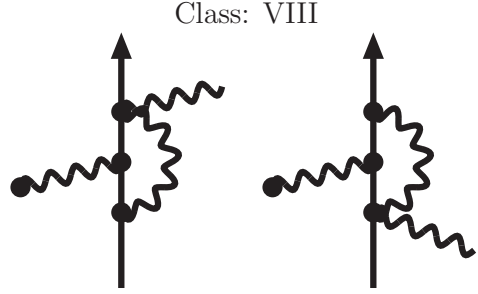}
\end{center}
\vspace{-.8cm}
\caption{One-photon loop diagrams of class VIII.}
\end{figure}

\noindent
\underline{Class VIII}: These irreducible diagrams give:
\begin{equation} A(s,u)^{(\rm VIII)} = {\alpha \over 2\pi} \int_0^1\!\!dx\!
\int_0^1\!\!dy\,\, {(1 -y)(2-y+x y) \over K(s,Q)}\,,\end{equation}
\begin{equation} B(s,u)^{(\rm VIII)} = {\alpha \over 2\pi} \int_0^1\!\!dx
\int_0^1\!\!dy\,\, {xy(y-1)\over K(u,Q)}\,. \end{equation}
The (singular) integral in eq.(14) involving $1/K(s,Q)$ requires a precise 
prescription for its (numerical) treatment. First, $s>1$ is given an 
infinitesimally positive imaginary part $s +i\,0^+$, which makes the integral 
complex-valued. In order to compute the only relevant real part of it, one 
performs analytically the Feynman parameter integral $\int_0^1dy$ and converts 
the occurring logarithm $\ln[x(1-s)]$ into $\ln[x(s-1)]$. The remaining 
integral $\int_0^1dx$ is free of poles and therefore unproblematic for a 
numerical treatment.   

\begin{figure}
\begin{center}
\includegraphics[scale=0.95,clip]{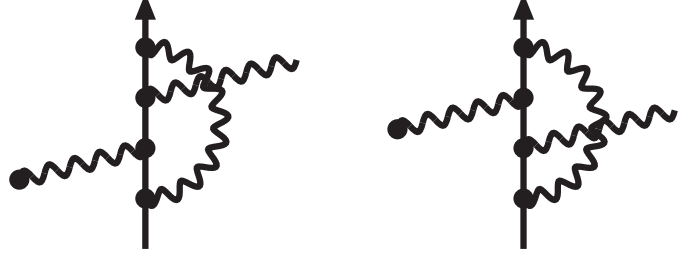}
\end{center}
\vspace{-.8cm}
\caption{One-photon loop diagrams of class IX.}
\end{figure}

\noindent
\underline{Class IX}: These box diagrams are most tedious to evaluate and
they give:
\begin{eqnarray} A(s,u)^{(\rm IX)} &=& {\alpha \over 2\pi} \Bigg\{{1\over u-1}- 
{u+1\over 2u(u-1)}\ln(1-u)+{1 \over (u-1)^2}\bigg[{\pi^2 \over 6}-{\rm  Li}_2(u)
\bigg] \nonumber \\ &&+{1\over (Q+t)^2}\bigg[ Q+t-4t\, L^2(-t) -4Q\, L^2(Q)+2 Q 
\sqrt{4-t} \, L(-t)\nonumber \\ && -2 Q \sqrt{4+Q} \, L(Q) \bigg]+ 
\int_0^1\!\!dx\!\int_0^1\!\!dy\,\bigg[{(1-x)y(1-y)\over K(s,Q)} +(2t-4)x y
\nonumber\\ && \times \int_0^1\!\!dz\,\bigg({(1-x)(1+xy-y -xy z)\over  
G^2(s,t,Q)}+{x(1-xy)(1-z)\over G^2(u,t,Q)}\bigg)\bigg]\Bigg\}\,,\end{eqnarray} 
where we have introduced the frequently occurring logarithmic loop function:
\begin{equation} L(x) = {1\over \sqrt{x}} \ln{\sqrt{4+x}+\sqrt{x}\over 2}\,,
\end{equation} 
and the quartic polynomial $G(s,t,Q)= y+(1-s)x(1-y)z-t x (1-x)y(1-z)+Q x(1-x)
y z$ in three Feynman parameters $x,y,z$. The other contribution reads:
\begin{eqnarray} B(s,u)^{(\rm IX)} &=& {\alpha \over 2\pi(u-1)} \Bigg\{{2u^2+u-1
\over 2u^2}\ln(1-u)-{u+1 \over 2u} +{1\over u-1}\bigg( {\rm  Li}_2(u)-{\pi^2 
\over 6}\bigg) \nonumber \\ && +{2t-4 \over\sqrt{t^2-4t}}\bigg[ 
4\Big( \xi_{IR}+\ln(1-u)\Big) \sqrt{-t}\, L(-t) + {\rm Li}_2(w)- {\rm Li}_2(1-w)
+{1\over 2} \ln^2 w \nonumber \\ &&- {1\over 2} \ln^2(1-w)+ {\rm Li}_2(h_-)- 
{\rm Li}_2(h_+) -\sqrt{t^2-4t}\int_0^1\!\!dx\,{\ln[1+Q x(1-x)]
\over 1-t x(1-x)}  \bigg]\  \nonumber \\ && + {u-1\over (Q+t)^2}\bigg[-Q-t
+4t\, L^2(-t) +4Q\, L^2(Q)-2 Q \sqrt{4-t} \, L(-t)\nonumber \\ && +2 Q \sqrt{
4+Q} \, L(Q) \bigg]+ (u-1)\int_0^1\!\!dx\!\int_0^1\!\!dy\,\bigg[{(1-xy)(y-1)
\over K(u,Q)}  \nonumber\\ &&+(2t-4) xy\int_0^1\!\!dz\,\bigg({x(1-x)y(z-1)
\over   G^2(s,t,Q)}+{1-xy+x^2y+xz- x^2yz \over G^2(u,t,Q)}\bigg)\bigg]\Bigg\}\,,
\end{eqnarray}
with the auxiliary variables: 
\begin{equation} w ={1\over 2}\Bigg(1-\sqrt{{-t \over 4-t}}\,\Bigg) \,, \qquad
h_\pm= {1\over 2} \Big( t \pm \sqrt{t^2-4t}\Big) \,,\end{equation}
depending on $t=2-s-u-Q$. In the physical region $u<1, t<0, Q>0$ the 
denominator polynomial $G(u,t,Q)>0$ is strictly positive and therefore the 
integrals involving $1/G^2(u,t,Q)$ are well-defined and suited for numerical 
evaluation. The (highly singular) integrals involving $1/G^2(s,t,Q)$ with 
$s>1$ (after the substitution $s \to s +i\,0^+$ which shifts the double-pole 
off the real integration contour) must however be treated partly analytically. 
The (solvable) integrals $\int_0^1  dz$ over the third Feynman parameter $z$ 
appear in the following two versions:
\begin{equation} \int_0^1\!\!dz\, {y \over G^2(s,t,Q)} = {1\over [1-t x(1-x)]
K(s,Q)}\,, \end{equation}
\begin{equation} \int_0^1\!\!dz\, {x^2z \over G^2(s,t,Q)} = {x \over N(s,Q+t)
K(s,Q)}+ {1\over N^2(s,Q+t)} \ln {K(s,Q) \over [1-t x(1-x)]y}\,, \end{equation}
giving rise to a new denominator polynomial $N(s,Q+t)=(s-1)(1-y)-(Q+t)(1-x)y
>0$ which is strictly positive for $s>1$ and $t< -Q$. Although the latter 
condition $t<-Q$ does not hold in the entire physical region, it represents
(fortunately) no restriction for the (experimentally selected) events in the 
Coulomb peak, where  $Q<0.1$ (or even less). By analyzing the condition $t<-Q$ 
in the laboratory frame one deduces the inequality for the ratio of photon to 
pion beam  energy, $\omega/E_1>[\sqrt{Q(4+Q)}-Q]/2 \simeq\sqrt{Q}-Q/2$, which is 
indeed well satisfied for the experimentally selected events. In the next step, 
the terms in eqs.(20,21) with a factor $1/K(s,Q)$ are integrated once more 
analytically, $\int_0^1 dy$, and for isolating the only relevant real parts the 
emerging logarithms are converted into logarithms of absolute values. The same 
conversion into real-valued logarithms is done with the term $\ln K(s,Q)$ in 
eq.(21), this time without solving the integral $\int_0^1 dy$ analytically. 
The remaining single and double integrals are free of poles and thus suitable 
for a straightforward numerical evaluation. The same strategy for evaluating the
occurring Feynman parameter integrals applies to the crossed amplitudes, 
$A(u,s)$ and $B(u,s)$, with the variables $s$ and $u$ interchanged. An 
alternative to the Feynman parameter representation of complicated loop 
functions in eqs.(11,14-18) is the (dispersive) spectral function 
representation based on the Kramers-Kronig dispersion relation (see 
section 3 in ref.\cite{radcor}). However, at non-zero $Q$ the corresponding 
spectral functions (or imaginary parts) entering these principal value 
integrals turn out to be very cumbersome. If an expansion up to linear order in
$Q$ suffices, one can do further steps in the analytical integration.

As a good check one verifies that the ultraviolet divergent terms proportional 
to $\xi_{UV}$ cancel in the total sum for $B(s,u)$. Interestingly, the 
contributions  to the other amplitude $A(s,u)$ written in eqs.(12,14,16) are
ultraviolet and infrared finite in each case.  
  
\section{Infrared finiteness}
In the next step we have to consider the infrared divergent terms proportional 
to $\xi_{IR}$. Inspection of eqs.(6,18) reveals that they scale with the (tree
level) Born amplitude written in eq.(4). As a consequence of that feature, 
the infrared divergent virtual (loop) corrections multiply the point-like
differential cross section $d\sigma^{(\rm  pt)}/d\omega d\Omega_\gamma d\Omega_\pi$ 
by a ($s \leftrightarrow u$ crossing-symmetric)  factor: 
\begin{equation} \delta_{\rm virt}^{(\rm IR)}= {2 \alpha \over \pi} \Bigg[ 1+
{2 t -4 \over \sqrt{4-t}} L(-t) \Bigg] \, \xi_{IR} \,. \end{equation}
The (unphysical) infrared divergence $\xi_{IR}$ gets canceled at the level of 
the (measurable) cross section by the contributions of soft photon 
bremsstrahlung. In its final effect, the (single) soft photon radiation
multiplies the tree level cross section $d\sigma^{(\rm pt)}/d\omega d\Omega_\gamma 
d\Omega_\pi$ by a (universal) factor \cite{radcor,radiat}: 
\begin{equation} \delta_{\rm soft}= \alpha\, \mu^{4-d}\!\!\int\limits_{|\vec 
l\,|<\lambda} \!\!{d^{d-1}l  \over (2\pi)^{d-2}\, l_0} \bigg\{ {m_\pi^2(2 - t) 
\over p_1 \cdot l \, p_2 \cdot l} - {m_\pi^2 \over (p_1 \cdot l)^2} - {m_\pi^2 
\over (p_2 \cdot l)^2} \bigg\} \,, \end{equation}
which depends on a small energy cut-off $\lambda$. Working out this
momentum space integral by the method of dimensional regularization (with 
$d>4$) one finds that the infrared divergent correction factor $\delta_{\rm  virt}
^{(\rm IR)} \sim \xi_{IR}$ in eq.(22) gets eliminated and the following finite 
radiative  correction factor remains:
\begin{eqnarray}\delta^{\rm (cm)}_{\rm real} &=&{\alpha\over\pi}\Bigg\{\bigg[2+{4t-8 
\over \sqrt{4-t}} L(-t)\bigg] \ln{m_\pi \over 2\lambda}+{s+1\over 2(s-1)}\ln s 
\nonumber \\ && +{s+1+Q \over \sqrt{(s-1+Q)^2 +4Q}}\ln{s+1+Q +\sqrt{(s-1+Q)^2 
+4Q} \over 2 \sqrt{s}}  \nonumber \\ &&+ (t-2) \int_0^1\!\!dx \,{ s+1+Q x
\over 2W[ 1-t x(1-x)]}\ln{ s+1+Q x+W\over s+1+Q x -W} \Bigg\}\,,\end{eqnarray} 
with the abbreviation $W = \sqrt{(s+1+Q x)^2-4s+4s t x(1-x)}$. We note that
the terms beyond those proportional to $\ln(m_\pi /2\lambda)$ are specific for 
the evaluation of the soft photon correction factor $\delta_{\rm soft}$ in the 
$\pi^-\gamma^*$ center-of-mass frame with $\lambda$ an infrared cut-off therein.
As an aside we mention that at $Q=0$ and in backward direction, $t=-(s-1)^2/s$, 
the integral in eq.(24) can be solved and one obtains the following handy 
formula:
\begin{eqnarray}\delta^{\rm (back)}_{\rm real} &=& {\alpha \over \pi (s^2-1)} \bigg\{ 
2\Big[s^2-1-(s^2+1)\ln s\Big]\ln{m_\pi \over 2 \lambda} \nonumber \\&& +(s+1)^2
\ln s +(s^2+1) \Bigg[ {1\over 2 }\ln^2 s +2 {\rm Li}_2(1-s) \bigg] \bigg\}\,.
\end{eqnarray}

\begin{figure}
\begin{center}
\includegraphics[scale=.47,clip]{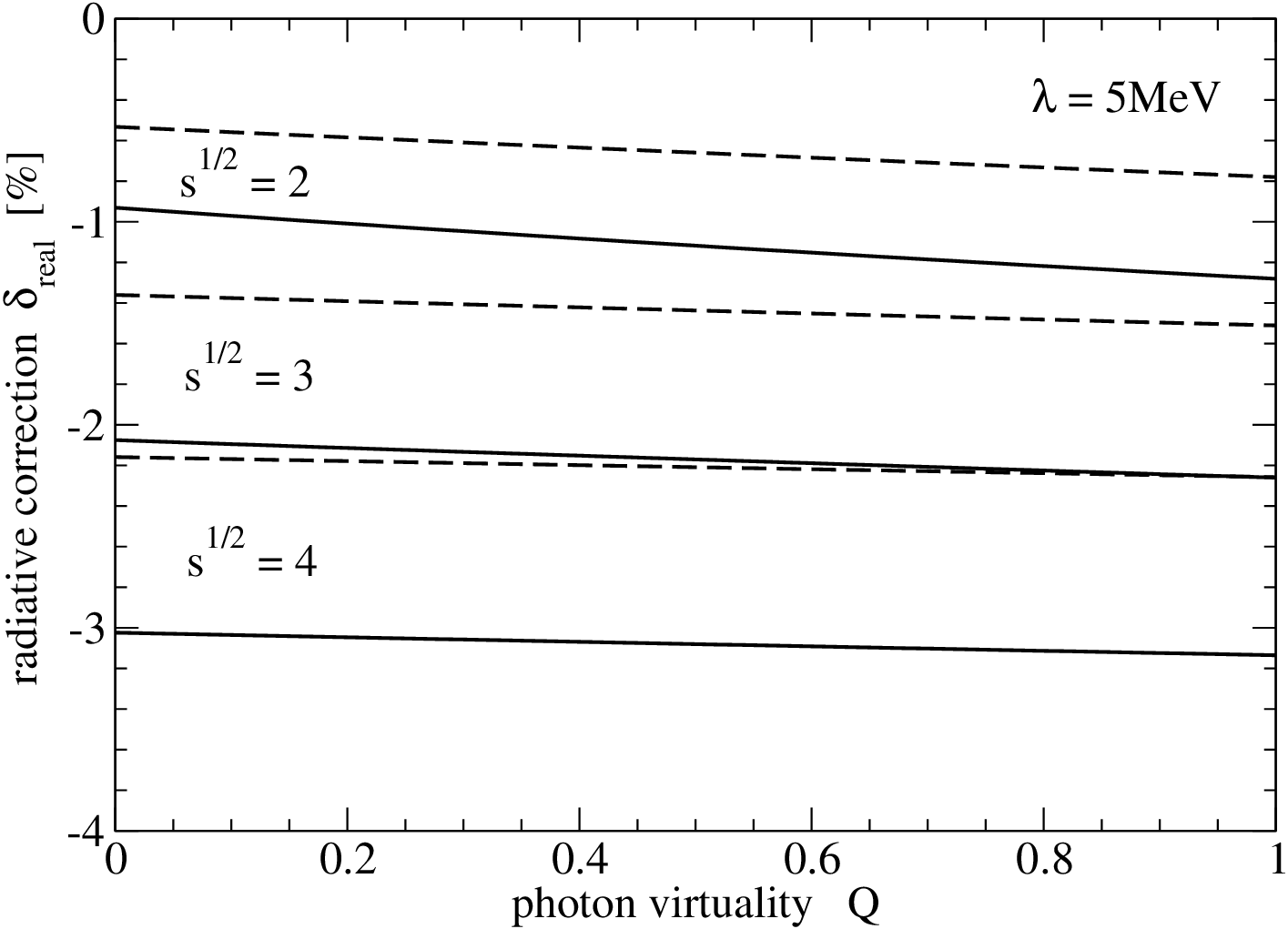}
\end{center}
\vspace{-.8cm}
\caption{Radiative correction factor $\delta_{\rm real}^{(\rm cm)}$ for virtual pion
Compton scattering $\pi^- \gamma^* \to \pi^- \gamma$ as a function of the
photon virtuality $Q m_\pi^2$. The soft photon detection threshold has been set 
to $\lambda =5\,$MeV in the $\pi^- \gamma$ center-of-mass frame. Full and
dashed lines correspond to backward and perpendicular scattering, respectively.}
\end{figure}

We can now present some numerical results for the radiative corrections to 
the pion-nucleus bremsstrahlung reaction $\pi^- Z \to \pi^- Z\gamma$. The 
radiative correction factor $\delta^{\rm (cm)}_{\rm real}$ due to soft photon emission
vanishes for $t=Q=0$ at all $s>1$. The full curves in Fig.\,11 show its values 
for three selected $\pi^-\gamma$ center-of-mass energies $\sqrt{s}m_\pi=(2,3,4)
m_\pi$ as a function of $Q$ (the photon virtuality divided by $m_\pi^2$) choosing 
the extremal momentum transfer $t_{\rm min}$, where $-2s\,t_{\rm min}=(s-1)^2+Q(s+1)
+(s-1)\sqrt{(s+1+Q)^2- 4s}$ corresponds to the backward direction in the $\pi^-
\gamma^*$ center-of-mass frame. The soft photon detection threshold has been 
set to $\lambda =5\,$MeV. One observes negative values up to $-3\%$ with a
weak dependence on $Q$ (in particular in the region of the Coulomb peak
$Q<0.1$). The three dashed lines in Fig.\,11 show for comparison the case of 
perpendicular scattering with momentum transfer $t_{\perp}$ reduced to about
one half of $t_{\rm min}$, where $-2s\,t_{\perp}=(s-1)^2+Q(s+1)$. At higher 
center-of-mass energies the excitation of the broad $\rho(770)$-resonance 
becomes prominent, $\sqrt{s_\rho} = m_\rho/m_\pi \simeq 5.5$. However, this 
kinematical region is not of interest for the extraction of the pion 
polarizabilities. The numbers in Fig.\,11 should be considered only as 
indicative since there are yet the (finite) virtual radiative corrections due 
to the photon-loops. A comparison with the case of real pion Compton
scattering \cite{radcor} suggests that these will reduce the magnitude of the 
radiative corrections to pion-nucleus 
bremsstrahlung. At the present stage, a proper quantification of these (finite) 
virtual radiative corrections is not possible since that requires specification 
of the actual experimental conditions, such as pion beam energy, constraints 
on detectable energy and angular ranges, etc. However, their full
implementation into the analysis of the COMPASS data is currently planned.     

\section{Inclusion of pion structure via polarizability difference}
In the same way as Akhundov et al. \cite{akhundov}, we have treated so far the
pion as a structureless spin-0 boson in our calculation of the radiative 
corrections. Now we go beyond this approximation which is valid only near the 
$\pi^-\gamma$ threshold. The leading pion structure relevant for (virtual) 
Compton scattering is given by the difference of its electric and magnetic 
polarizability $\alpha_\pi- \beta_\pi$. In an effective field theory approach
it is easily accounted for by a new two-photon contact-vertex proportional to 
the squared  electromagnetic field strength tensor, $F_{\mu\nu}F^{\mu\nu} = 2(\vec
B^2- \vec E^2)$. The S-matrix insertion following from this higher-order 
(gauge-invariant) effective $\pi\pi\gamma\gamma$-vertex reads: 
\begin{equation} 8\pi i \beta_\pi m_\pi \Big( k_1 \cdot k_2 \, \epsilon_1  \cdot 
\epsilon^*_2 -\epsilon_1\cdot k_2\,\epsilon^*_2\cdot k_1\Big)\,,\end{equation} 
where one photon $(k_1, \epsilon_1)$ is ingoing and the other one $(k_2, 
\epsilon_2)$ outgoing. At tree level (see left diagram in Fig.\,12) the 
polarizability vertex eq.(26) gives in the kinematical situation of $\pi^- 
\gamma^*_0 \to \pi^- \gamma$ rise to the (constant) contributions:
\begin{equation} A(s,u)^{(\rm pola)} =- B(s,u)^{(\rm pola)} =-\,{\beta_\pi m_\pi^3 \over  
\alpha} \,,\end{equation}
to the amplitudes $A(s,u)$ and $B(s,u)$ parameterizing the laboratory T-matrix 
$T_{\rm lab}$. In order to prevent any misunderstandings we stress that when 
writing (merely) $\beta_\pi m_\pi$ for the coupling strength in eq.(26), an 
electric and a magnetic pion polarizability, equal in magnitude and opposite 
in sign $\alpha_\pi = -\beta_\pi$, are always both included.

We reinterpret  now the one-photon loop diagrams of section 3 in such a way 
that the two-photon contact-vertex represents one polarizability vertex 
proportional to $\beta_\pi m_\pi$. In order to distinguish it from the leading 
order contact-vertex $8\pi i\alpha\, \epsilon_1\cdot \epsilon^*_2$ of scalar
QED we have symbolized it in Figs.\,12-15 by an open square. Going through the 
classes I'- V' and evaluating the loop diagrams with the S-matrix insertion
from the polarizability vertex we find the following contributions to the
amplitudes  $A(s,u)$ and $B(s,u)$.

\begin{figure}
\begin{center}
\includegraphics[scale=0.95,clip]{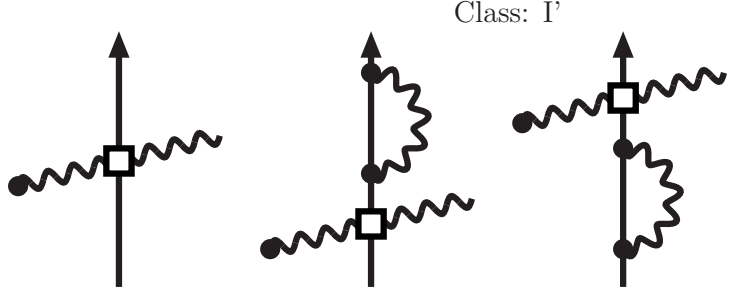}
\end{center}
\vspace{-.8cm}
\caption{The open square symbolizes the electromagnetic interaction
proportional to the (dominant) pion polarizability difference $\alpha_\pi-
\beta_\pi$. Shown are the tree diagram and the diagrams of class I' involving
the pion wavefunction renormalization factor.}
\end{figure}

\noindent
\underline{Class I'}:
\begin{equation}  A(s,u)^{(\rm I')} =- B(s,u)^{(\rm I)} = {\beta_\pi  m_\pi^3 \over    
\pi} \Big( \xi_{UV}  -\xi_{IR} \Big)\,.\end{equation}

\begin{figure}
\begin{center}
\includegraphics[scale=0.95,clip]{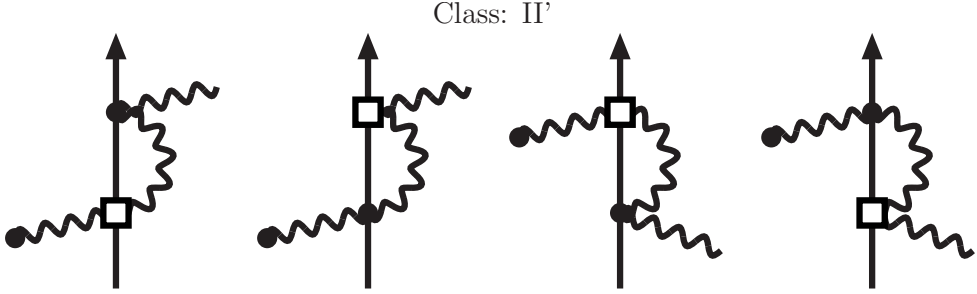}
\end{center}
\vspace{-.8cm}
\caption{One-photon loop diagrams of class II' including pion structure via
its polarizability difference.}
\end{figure}

\noindent
\underline{Class II'}:
\begin{equation} A(s,u)^{(\rm II')} = {\beta_\pi  m_\pi^3 \over 2 \pi}\bigg[-2  
\xi_{UV} +2 -{1\over 2s}-{1\over 2u}-{(s-1)^2 \over 2s^2}\ln(1-s)-{(u-1)^2
\over 2u^2} \ln(1-u)\bigg]\,,\end{equation}
\begin{equation} B(s,u)^{(\rm II')} =  {\beta_\pi  m_\pi^3 \over 2 \pi}\bigg[2  
\xi_{UV} -2 +{1\over u}+{(u-1)^2 \over u^2}\ln(1-u)\bigg]\,.\end{equation}

\begin{figure}
\begin{center}
\includegraphics[scale=0.95,clip]{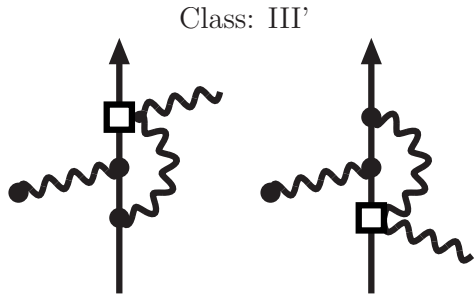}
\end{center}
\vspace{-.8cm}
\caption{One-photon loop diagrams of class III'.}
\end{figure}

\noindent
\underline{Class III'}
\begin{equation} A(s,u)^{(\rm III')}= { \beta_\pi  m_\pi^3 \over 2 \pi}\bigg\{-\xi_{UV}
+{1\over4} + \int_0^1\!\!dx\!\int_0^1\!\!dy\,y \bigg[{(s-1)x(1 -y) \over K(s,Q)}
- \ln K(s,Q) \bigg] \bigg\}\,,\end{equation}
\begin{equation} B(s,u)^{(\rm III')} = { \beta_\pi  m_\pi^3\over 2\pi} \bigg\{ \xi_{UV}
-{1\over4} + \int_0^1\!\!dx\!\int_0^1\!\!dy\,y \bigg[{(s-1)x(1 -y) \over K(u,Q)}
+ \ln K(s,Q) \bigg] \bigg\}\,,\end{equation}
with the polynomial $K(u,Q)$ defined below eq.(11). The strategy to compute
the integrals involving $1/K(s,Q)$ and $\ln K(s,Q)$ has been described in
section 3.  
\begin{figure}
\begin{center}
\includegraphics[scale=0.95,clip]{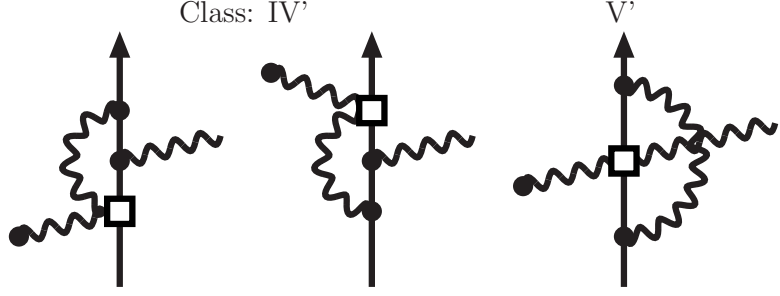}
\end{center}
\vspace{-.8cm}
\caption{One-photon loop diagrams of classes IV' and V'.}
\end{figure}

\noindent
\underline{Class IV'}:
\begin{eqnarray} A(s,u)^{(\rm IV')} &=& {\beta_\pi m_\pi^3\over 2 \pi}\bigg\{-\xi_{UV} 
+{3\over 2} -{u+1 \over 2u} \ln(1-u) + {1 \over u-1}\bigg[{\pi^2 \over 6}-{\rm 
Li}_2(u)\bigg]   \nonumber \\ && +{u-1-Q \over u-1} \bigg[ {3u-1 \over 2u^2} 
\ln(1-u)-  {u+1 \over 2u} +{1 \over u-1}\bigg({\rm Li}_2(u)-{\pi^2 \over 6}
\bigg) \bigg] \bigg\} \,, \end{eqnarray}
\begin{eqnarray} B(s,u)^{(\rm IV')} &=& { \beta_\pi  m_\pi^3\over 2\pi}\bigg\{\xi_{UV} 
-{3\over 2} +{s+1 \over 2s} \ln(1-s) + {1 \over s-1}\bigg[{\rm Li}_2(s)-{\pi^2 
\over 6}\bigg]   \nonumber \\ && +{s-1+Q \over u-1} \bigg[ {3u-1 \over 2u^2} 
\ln(1-u)-  {u+1 \over 2u} +{1 \over u-1}\bigg({\rm Li}_2(u)-{\pi^2 \over 6}
\bigg) \bigg] \bigg\} \,. \end{eqnarray}

\noindent
\underline{Class V'}:
\begin{eqnarray} A(s,u)^{(\rm V')}&=&-B(s,u)^{(\rm V')}={\beta_\pi m_\pi^3 \over 2\pi}
\bigg\{\xi_{UV} -1- \sqrt{4 -t}\, L(-t) + {2-t \over  \sqrt{t^2-4t}}\nonumber
\\ && \times \bigg[4 \xi_{IR} \sqrt{-t}\,L(-t)+{\rm  Li}_2(w)-{\rm  Li}_2(1-w) 
+{1\over 2} \ln^2 w -{1\over 2}  \ln^2(1-w) \bigg] \bigg\}\,, \end{eqnarray}
with the auxiliary variable $w$ defined in eq.(19). The contributions of the
vertex type diagrams (classes III and IV in Figs.\,4,5) vanish due to some
special features of the polarizability vertex eq.(26).  

At first count the ultraviolet divergent terms proportional to $\xi_{UV}$ do not
drop out in the total sums for $A(s,u)$ and $B(s,u)$. They disappear however
after interpreting the coupling constant in eq.(27) as a bare one and 
splitting it as $\beta_\pi^{(\rm bare)} = \beta_\pi (1 - \alpha\,\xi_{UV}/2\pi)$ into 
the physical coupling constant $\beta_\pi$ and a counterterm piece.
Note that the same renormalization procedure has rendered the radiative 
corrections to real pion Compton scattering ultraviolet finite \cite{radcor}. 
The infrared divergent terms proportional to $\xi_{IR}$ showing up in 
eqs.(28,35) are again canceled by the soft photon bremsstrahlung 
contributions. As a result one has to multiply the partial differential cross 
section $d\sigma^{(\rm pol)}/d\omega d\Omega_\gamma d\Omega_\pi$ linear in 
$\beta_\pi$, which arises from the interference of the Born and polarizability 
terms in eqs.(4,27), with the radiative correction factor  
$\delta^{(\rm cm)}_{\rm real}$ written down in eq.(24). In the case of real 
pion Compton scattering it has been observed in ref.\cite{radcor} that the 
inclusion the leading pion structure effect (via $\alpha_\pi -\beta_\pi \simeq 
-2\beta_\pi$) does not change the relative size and angular dependence of the 
radiative corrections. Due to the dominance of real over the virtual radiative 
corrections (for small enough infrared cut-off $\lambda$) one can expect that 
the same feature holds for the pion-nucleus bremsstrahlung reaction 
$\pi^- Z \to \pi^- Z\gamma$.

\section*{Appendix: Specific two-photon exchange contributions}
For high nuclear charges $Z$ the one-photon exchange approximation is 
insufficient for analyzing experimental data of pion-nucleus bremsstrahlung 
$\pi^- Z\to\pi^-Z\gamma$. A non-perturbative treatment of the scattering of the
charged particles in the strong nuclear Coulomb field is required in this 
case. Integrating the tree-level cross section  $d\sigma^{(\rm  pt)}/d\omega
d\Omega_\gamma d\Omega_\pi$  of scalar QED over the angles $(\theta_1,
\theta_2,\phi)$, and adapting the result of refs.\cite{bethe,landau} 
obtained for electrons to pions, one may write the photon spectrum at high beam 
energies $E_1$ in the form:
\begin{equation} {d \sigma \over d\omega}^{(\rm  pt)} = {16Z^2\alpha^3E_2 \over 
3m_\pi^2 E_1 \omega} \bigg\{ \ln{2E_1 E_2 \over m_\pi\omega}-{1\over 2} 
+\Psi_\pi(Z\alpha)\bigg\}\,, \end{equation}
with $\omega= E_1-E_2$ the photon laboratory energy. The function $\Psi_\pi(Z
\alpha)$ represents the non-perturbative corrections beyond the one-photon 
exchange. For electrons this function has been calculated as $\Psi_e(Z\alpha) = 
-Z^2 \alpha^2 \sum_{n=1}^\infty [n(n^2+Z^2 \alpha^2)]^{-1}$ in 
refs.\cite{bethe,landau}. In the case of a lead or a nickel target ($Z=82$ or 
$28$) the negative infinite series $\Psi_e(Z\alpha)$ amounts to $-0.3316$ or 
$-0.0484$, i.e. about a $-5\%$ or $-0.7\%$ correction compared to the leading 
logarithm plus constant in eq.(36). At this point it would be most important to
clarify, whether these non-perturbative corrections are the same for pions and
electrons or not.   

\begin{figure}
\begin{center}
\includegraphics[scale=0.95,clip]{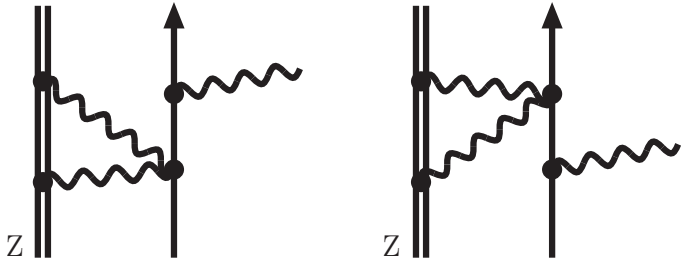}
\end{center}
\vspace{-.8cm}
\caption{Two-photon exchange diagrams specific for the contact-vertex of scalar 
QED. The double-line symbolizes the heavy nucleus of charge $Z$.} 
\end{figure} 

As a supplement to the subject, we evaluate in this appendix those two-photon 
exchange contributions (of order $Z^2$ in the transition amplitude) which are 
specific to the two-photon contact-vertex of scalar QED. The pertinent one-loop 
diagrams are shown in Fig.\,16. Additionally possible box and crossed box type 
diagrams, where one exchanged virtual photon and the emitted real photon are
connected by the contact-vertex vanish as a consequence of the gauge condition 
$\epsilon_0=0$. For the evaluation of the (triangular) loop we use a 
non-relativistic (static) propagator $i/(l_0+i\,0^+)$ for the intermediate 
heavy nucleus. By dividing out the factor $-e Z/|\vec q\,|^2$ we can translate 
the two-photon exchange processes into a correction to the T-matrix $T_{\rm lab}$ 
of virtual pion Compton scattering. Putting all pieces together we find from 
the photon-loop diagrams in Fig.\,16 the following contribution: 
\begin{equation} T_{\rm lab}^{(2\gamma)} =(2\pi \alpha)^2 Z {\sqrt{Q} \over m_\pi}
\Bigg( {\vec \epsilon^{\,\,*}\cdot \vec p_1 \over u-1} + {\vec  \epsilon^{\,\,*}
\cdot \vec p_2 \over s-1} \Bigg) \,. \end{equation}
It has a structure very similar to the T-matrix in Born approximation (see 
eqs.(3,4)). Making this comparison explicitly, one deduces that the coupling 
terms $\vec \epsilon^{\,\,*}\cdot \vec p_{1,2}$ in eq.(37) carry the relative 
suppression factors $\pi \alpha Z m_\pi\sqrt{Q}/4E_{2,1}$, respectively. Although 
$\pi \alpha Z/4\simeq 0.47$ is already sizeable for a lead target ($Z=82$), 
the ratio $|\vec q\,|/E_{1,2}$ of (virtual photon) momentum transfer to beam or 
scattered pion energy reduces it by at least three orders of magnitude. We are
considering here the typical kinematical situation, $E_{1,2}> 50\,$GeV and 
$|\vec q\,|<50\,$MeV, where high-energy pion-nucleus bremsstrahlung $\pi^- Z 
\to \pi^- Z \gamma$ is used to extract the pion polarizabilities. In passing
we note that the diagrams in Fig.\,16 with the contact-vertex replaced by the
polarizability vertex give rise to a T-matrix that differs from eq.(37) merely
by an additional prefactor $\beta_\pi m_\pi^3  Q/2\alpha \simeq -7\cdot 10^{-3}\,
Q$ (using $\alpha_\pi = - \beta_\pi \simeq  3 \cdot 10^{-4}\,$fm$^3$). We can 
therefore conclude that the two-photon exchange mechanisms in Fig.\,16,
specific to the pseudo-scalar pion, are negligibly small.

\end{document}